\documentclass[10pt,preprint]{aastex}

\newcommand{\msun}{\,M$_\odot$}

\def\deg{$^{\circ}$}
\def\solm{M$_{\odot}$}
\def\kms{km\,s$^{-1}$}

\def\d{{\rm d}}

\shorttitle{Spiral Driven Inflow in NGC1097}
\shortauthors{Davies et al.}

\begin{document}


\title{Stellar and Molecular Gas Kinematics of NGC1097: Inflow Driven
  by a Nuclear Spiral\footnote{Based on observations at the ESO Very
    Large Telescope (076.B-0098)}}


\author{R.I. Davies}
\affil{Max Planck Institut f\"ur extraterrestrische Physik, Postfach 1312,
  85741, Garching, Germany}
\email{davies@mpe.mpg.de}

\and

\author{W. Maciejewski}
\affil{Astrophysics Research Institute, Liverpool John Moores University, Twelve Quays House, Egerton Wharf, Birkenhead CH41 1LD, UK}

\and

\author{E.K.S. Hicks, L.J. Tacconi, R. Genzel, H. Engel}
\affil{Max Planck Institut f\"ur extraterrestrische Physik, Postfach 1312,
  85741, Garching, Germany}



\begin{abstract}
We present spatially resolved distributions and kinematics of the
stars and molecular gas in the central 320\,pc of NGC1097.
The stellar continuum confirms the previously reported 3-arm spiral
pattern extending into the central 100\,pc.
The stellar kinematics and the gas distribution
imply this is a shadowing effect due to extinction by gas and dust
in the molecular spiral arms.
The molecular gas kinematics show a strong residual
(i.e. non-circular) velocity, which is manifested as a 2-arm
kinematic spiral. Linear models indicate that this is the line-of-sight
velocity pattern expected for a density wave in gas that generates a 3-arm
spiral morphology.
We estimate the inflow rate along the arms.
Using hydrodynamical models of nuclear spirals, we show that when
deriving the accretion rate into the central region, outflow in the
disk plane between the arms has to be taken into account.
For NGC\,1097, despite the inflow rate along the arms being
$\sim1.2$\msun\,yr$^{-1}$, the {\em net} gas accretion rate to the
central few tens of parsecs is much smaller. The numerical models
indicate that the inflow rate could be as little as
$\sim0.06$\msun\,yr$^{-1}$. 
This is sufficient to generate recurring starbursts, similar in scale
to that observed, every 20--150\,Myr.
The nuclear spiral represents a mechanism that can feed gas into the
central parsecs of the galaxy, with the gas flow sustainable for
timescales of a Gigayear.
\end{abstract}





\section{Introduction}
\label{sec:intro}

During the past decades there has been significant and increasing
interest in mechanisms that can bring gas from remote locations in a
galaxy to the nucleus, where it could fuel the central black hole.
Such mechanisms would have to extract nearly all of the angular
momentum from the gas if it is to be brought from a
distance of a few kiloparsec to a few Schwarzschild radii from the central
black hole. No unique mechanism has been found to operate over this
entire range of radii, and the evidence is growing that no such
single mechanism exists \citep{shl90,mul97,mar03,hun04,wad04}.
In particular, it has been found that active star
formation takes place in the innermost few parsecs of galaxies
\citep{fab98,lev01,wad02,cid04,tho05,dav07,bal08}, which would
disturb any flow pattern established at larger scales.

On scales of kiloparsecs, the most efficient drivers of gas inflow involve
gravity torques, especially coming from passing companions, and
non-axisymmetries in the disk, like bars \citep{com03,kor04,gar05}.
Observations show that the
presence of companions correlates with star-forming activity in centres of
galaxies but not with activity related to
the central black hole \citep{sch01,li08a,li08b}.
A similar situation holds for bars, with Seyfert galaxies
showing at best only a marginal statistical excess of bars with respect to
non-Seyfert galaxies \citep{ho97,mul97,lai02,com03}.
\cite{shl00} even reported a slight deficiency of strong bars in
Seyfert galaxies.
Gas inflowing along the bar
often settles on star-forming nuclear ring, about 1\,kpc from the galaxy
centre, and inflow inside the ring is small \citep{pin95,reg03}.
The inflow may be
extended inwards in nested bars, as proposed by \cite{shl89}, but
dynamical constraints on nested bars may prohibit inflow
\citep{mac02}.
\cite{eng00} noticed that inside of strong inflow
in a bar, instead of a nuclear ring, a nuclear spiral may form. Inflow in
the bar along straight shocks curves towards the centre and turns into a
spiral pattern. In the models of \cite{eng00} this gaseous nuclear
spiral is no longer a shock, but a wave well described by the linear
density wave theory. To the contrary, hydrodynamical models by
\cite{mac04b} showed that the nuclear spiral can propagate to the centre
of a galaxy as a shock in gas. Strong streaming motions (i.e. large
velocity residuals), which take the form of kinematic spiral arms, are
expected in the nuclear spiral shock \citep{mac06}. Kinematic spiral arms
were already observed in emission from ionized gas in NGC\,1097 \citep{fat06},
NGC\,6951 \citep{sto07} and M\,83 \citep{fat08}.

In this paper, we examine this mechanism with reference to NGC\,1097,
a nearby SBbc galaxy (D=18\,Mpc for
$H_0$=70\,km\,s$^{-1}$\,Mpc$^{-1}$; 1\arcsec$\sim85$\,pc). It
has a strong large-scale bar and a
800\,pc radius circumnuclear ring that is rich in molecular
gas \citep{ger88,koh03,hsi08} and vigorously forming stars
\citep{hum87,kot00}.
Inside this ring, a nuclear spiral extends down to the central few
tens of parsecs \citep{pri05,fat06}.
NGC\,1097 hosts a weak AGN, classified as both a LINER and, from
evidence of the broad line region \citep{sto93} and the hard X-ray
excess \citep{iyo96}, as a type~1 Seyfert.
Over the last decade or two, the AGN luminosity has gradually decreased
\citep{sto03}, and hydrogen recombination emission from
the nuclear region is currently very weak \citep{dav05,dav07}.
This is fortuitous, since it enables one to look very close in to the
nucleus without being blinded by the AGN itself.
We present SINFONI data for NGC\,1097 extending out to a radius of
160\,pc (see also \citealt{dav07,hic09}),
and analyse the distribution and kinematics of the stellar continuum
and molecular gas emission in the context of the nuclear spiral.

In Section~\ref{sec:obs} we summarise the observations and data
reduction, including the method used to extract kinematics and
morphologies from the spectral features, an estimation of the point
spread function (PSF), and
the adopted orientation of the galaxy.
We then analyse the stellar continuum in Section~\ref{sec:stars} and
the molecular gas in Section~\ref{sec:h2}, including an estimation of
the gas surface density in and between the spiral arms.
In Section~\ref{sec:models} we discuss the dynamics of nuclear
spirals, and then in Section~\ref{sec:transport} estimate the inflow
rate for NGC\,1097.
We discuss the implications of our results in Section~\ref{sec:disc}
before finally presenting our conclusions.

\section{Observations and Data Processing}
\label{sec:obs}

The data on which the present analysis of inflow is performed,
have been presented previously in \cite{dav07} and \cite{hic09}.
These papers focussed on the global properties of the stars and gas in
samples of nearby active galaxies.
\cite{dav07} (Section A2.4) showed that there was a kinematically
(i.e. lower dispersion) and photometrically (i.e. excess
stellar continuum) distinct stellar population in the nucleus, which had
the properties of a short-lived and still young (i.e. $<$10\,Myr old)
starburst; and that the stars and molecular gas must be mixed.
\cite{hic09} (Section A.1) analysed the molecular gas, showing that it
exhibits the 
characteristics expected for the large scale properties of the
obscuring molecular torus. Specifically, the increase in
the gas dispersion at radii less than 30\,pc is due to a thickening of
the gas disk; and using also millimetre CO measurements, that the gas
column density in this region is of order $10^{23}$\,cm$^{-2}$.

\subsection{Observations \& Reduction}

The observations of NGC\,1097 were performed on the night of
10-Oct-2005 at the VLT with SINFONI, an adaptive optics near infrared
integral field spectrograph \citep{eis03,bon04}.
Data were taken with the H+K grating (covering both bands
simultaneously at a resolution of $R\sim1500$), and with a pixel scale
of 0.05\arcsec$\times$0.1\arcsec.
Individual exposure times were 2$\times$300\,sec, and 6 pairs of
frames were taken in the sequence O-S-O-O-S-O, to facilitate
background subtraction, yielding a total on-source integration time of
40\,mins.
The data were processed using the dedicated {\it spred} software package
\citep{abu06}, which provides similar processing to that for long slit
data but with the added ability to reconstruct the datacube.
The data processing steps were as follows.
The object frames were pre-processed by subtracting sky frames,
flat-fielding, and correcting bad pixels (which are identified from
dark frames and the flatfield).
The wavemap was generated, and edges and curvature of the slitlets were
traced, all from the arclamp frame.
The arclamp frame was then reconstructed into a cube, which was checked
to ensure that the calibration is good.
The pre-processed object frames were then also reconstructed into
cubes.
At this stage the wavelength calibration was fine-tuned for each cube
separately with reference to the OH lines using a custom script.
And in order to optimize the subtraction of the OH lines, the cubes
were further processed using the method described in \cite{dav07a}.
Following this, the cubes were spatially aligned using the bright
nucleus as a reference, and finally combined.
We note that during this process, an unusual but minor affect was
imprinted into the data.
This consists of slight changes between adjacent slitlets, which
manifests themselves as small linear features in alternate rows, both
in the flux and kinematics maps.
Because the effect is small, and occurs at a spatial frequency
significantly higher than the resolution, it has no impact on the
analysis.
The pixel scale of the final cube was 0.05\arcsec$\times$0.05\arcsec.
Estimation of the spatial resolution (see below) was performed
following extraction of the stellar continuum properties.

The standard star frames were similarly reconstructed into cubes.
Telluric correction and flux calibration were performed with reference
to the B6V star HD\,28107.
Because it is not easily possible to correct for all the absorption
features in the H-band spectrum of such stars, the stellar spectrum
was used only to derive the instrumental transmission as a function of
wavelength.
This was then combined with a theoretical model of the atmospheric
transmission, which was adjusted to provide the best correction for
the data.
The zero-point derived from the standard star was
consistent with that expected based on other standard stars taken at
different times.
The flux calibration was nevertheless cross-checked in 3\arcsec\ apertures
using 2MASS data, and in smaller 1--3\arcsec\ apertures using
broad-band imaging from NACO.
The calibration of our data was consistent with the external data to
20-30\%.

\subsection{Emission/Absorption Line Characterisation}
\label{sec:linefit}

The 2D distribution and kinematics of emission lines was found
using the code LINEFIT developed by our group specifically for SINFONI
applications (Davies et al. in prep.; see also \citealt{for09}).
LINEFIT fits a function to the continuum-subtracted spectral profile at
each spatial position in the datacube.
The function is a convolution of a Gaussian with a spectrally
unresolved template profile -- an OH sky emission line.
A minimisation is performed in which the parameters of the Gaussian
are adjusted until the convolved profile best matches the data.
During the minimisation, pixels in the data that consistently
deviate strongly from the data are rejected.
The uncertainties are boot-strapped using Monte Carlo techniques.
Before extraction of the emission line properties, a 3$\times$3 pixel
median filter was used to smooth each spatial plane of the cube.
While this affects the spatial resolution, it has no impact on the
spectral properties.
The key features of LINEFIT include:
(i) the spectral
resolution is implicitly taken into account by convolving the
assumed emission line profile with a template line (tracing the
effective instrumental resolution) before performing the fits;
(ii) weighted fits are performed according to three possible
schemes -- in this instance we used uniform (i.e. no) weighting, since
the noise was set to be the rms of the fit to the continuum at each
spaxel;
(iii) formal
uncertainties are computed from 100 Monte Carlo simulations,
where the points of the input spectrum at each spatial
pixel are perturbed assuming Gaussian noise properties characterized
by the rms from the noise cube.

The 2D properties of the stellar absorption features are extracted
using an almost equivalent code.
The only differences are that the convolution is performed using
spectra of template stars observed independently but in the same
instrument configuration (pixel scale and grating);
and the uncertainties are derived assuming that the extracted profile
is well represented by a Gaussian (an assumption that is not made in
LINEFIT).
As for the emission line, each spatial plane was smoothed with a
3$\times$3 pixel median filter before performing the extraction.

During this procedure, the stellar and non-stellar continua are
separated using the method described in Section~3.1 of \cite{dav07}.
Basically, stellar population synthesis models, as well as
observations of galaxies, show that the equivalent width of
the 2.3\micron\ CO bandhead integrated over an evolving stellar
cluster is approximately independent of age (for ages greater than
10\,Myr).
Any significant downward deviations from this value can be
attributed to dilution by non-stellar continuum.
At our resolution, within the central 0.5\arcsec\ of NGC\,1097,
non-stellar emission comprises slightly over 50\% of the 2.3\micron\
continuum.
Within 1\arcsec\ and 2\arcsec\ apertures, this fraction drops to 40\%
and 35\% respectively.

\subsection{PSF Estimation}
\label{sec:psf}

There are a multitude of ways to derive the PSF from adaptive optics
data, five of which are described in \cite{dav07b}.
With AGN, it is usually possible to estimate the PSF from the science
data itself, removing any uncertainty about spatial and temporal
variations of the PSF due to atmospheric effects.
For NGC\,1097, the resolution was determined from the non-stellar
K-band continuum.
This will be unresolved in all but the nearest AGN since at near infrared
wavelengths it is expected to originate from a region no more than
1--2\,pc across.
The non-stellar continuum is one of the frames output during the
extraction of the stellar properties, and therefore already includes
the effects of the spatial smoothing that was applied to the cube.

We have fit an analytical function to the PSF.
Since the Strehl ratio achieved is relatively low,
even a Gaussian is a reasonable approximation.
However, we have used a Moffat function, which achieves a better fit
because it also matches the rather broad wings that are a characteristic of
partial adaptive optics correction.
If one applies the concept of `core plus halo' to this PSF, then the
Gaussian fit would represent just the core while the Moffat fits the
entire `core plus halo'.
Both functions yield a K-band FWHM of 0.25\arcsec.
Integrating both of these functions indicates that about 75\% of the
flux is within the `core', and it is thus this component which
dominates the PSF.

\subsection{Position Angle and Axis Ratio}
\label{sec:paaxrat}

The position angle and axis ratio for the central few hundred parsecs
of NGC\,1097 have been derived previously for these SINFONI data.
For each of these parameters, \cite{dav07} and \cite{hic09} determined
very similar values (i.e. to within a few degrees) from both the stellar
velocity field and the gas velocity field. These values also agree
well with those estimated by \cite{sto03} and \cite{fat06}.
We have adopted typical values here: a position angle of -49$^\circ$ and
an axis ratio of 0.85 (i.e. inclination of 32$^\circ$).

\section{Stellar Distribution and Kinematics}
\label{sec:stars}

\begin{figure}
\epsscale{.95}
\plotone{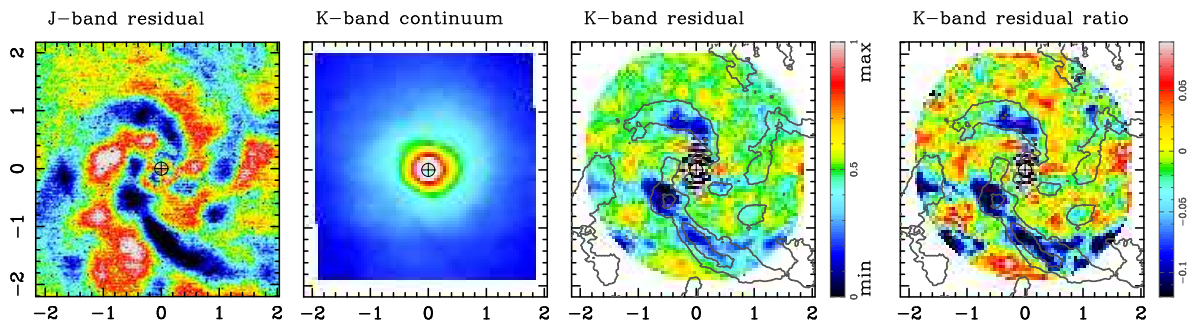}
\caption{Far Left: NACO J-band continuum residual (from, and as presented in,
  \citealt{pri05}) after subtracting elliptical isophotes.
Middle Left: SINFONI K-band continuum, showing the bright nucleus
  containing a significant fraction of non-stellar light.
Middle Right: the same K-band continuum after subtracting elliptical
  isophotes.
Far Right: ratio of the residual to the smooth isophotal model (from
  which the central non-stellar point source has been subtracted). 
  This allows one to read-off almost directly as 0.1--0.15 the contrast ratio
  described in Section~\ref{sec:gasden}.
Superimposed are contours outlining the negative J-band residual in the
  left-most panel.
The same 3-armed spiral structure is apparent in both J-band NACO data
  and SINFONI K-band data.
In all panels, north is up and east is left; and the scale in arcsec
  (with 1\arcsec\ = 85\,pc) is indicated.
\label{fig:starflux}
}
\end{figure}

\cite{pri05} showed that although the total J-band continuum increased
smoothly towards the nucleus, the residuals -- after subtracting
elliptical isophotes -- revealed a 3-arm spiral structure in absorption.
This 3-arm structure is confirmed by our SINFONI data.
In Fig.~\ref{fig:starflux} we show the K-band continuum distribution,
together with the residual after fitting and subtracting elliptical
isophotes. 
\cite{pri05} decomposed the near infrared continuum from the central
few arcsec into a contributions from a bulge and a point source.
\cite{dav07} confirm the presence of a bulge, measuring $V/\sigma<0.8$
out to 2\arcsec.
But in addition, both they and \cite{ems01} 
found evidence for a drop in dispersion close to the nucleus, indicating a
kinematically distinct stellar population within the central arcsec.
While our primary interest is in the
residuals rather than the fit itself, our fitting procedure needs to
be sufficiently representative.
As a result, in order to allow for multiple stellar components as well
as possible isophotal twists of the bulge, we have left the axis
ratio and the position angle, as well as the center, as free
parameters for each isophote.
These extra degrees of freedom, by making it possible to remove global
gradients that would otherwise have remained, yield
a more uniform residual with clearly defined structures.
There are three arms in the residual: the northeastern and the southwestern
arms are strong, and the third arm, towards the northwest, is weaker.
The overlay shows that they do trace the same pattern, although less
distinct, as in the J-band NACO data presented by \cite{pri05}.

\begin{figure}
\epsscale{.95}
\plotone{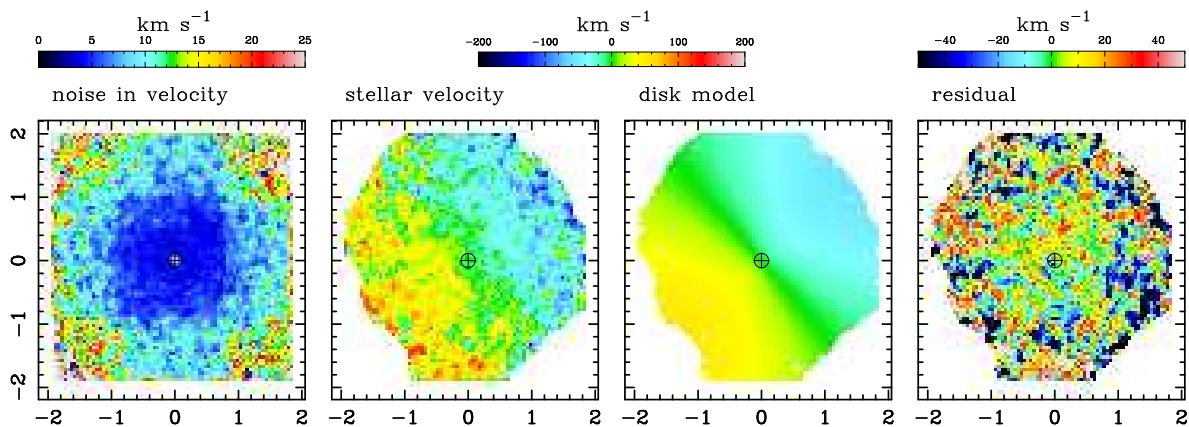}
\caption{Far left: noise in stellar velocity field. The pattern is
  dominated by the effects of dithering (i.e. the central part of the
  field with the longest integration time has uncertainties of
  5\,km\,s$^{-1}$ or less, and the corners have the least integration
  and highest errors).
Center left: the stellar velocity field derived by fitting spectral
  templates to the 2.3\micron\ CO bandhead. Regions of particularly
  high noise have been excised.
Center right: the best fitting axisymmetric disk model velocity field.
Far right: the residual after subtracting the model velocity field from
  that observed shows no coherent structures above the noise level.
In all panels, north is up and east is left; and the scale in arcsec
  (with 1\arcsec\ = 85\,pc) is indicated.
\label{fig:starvel}
}
\end{figure}

The SINFONI data have the advantage of probing also the kinematics,
and we present the stellar velocity field -- together with its
residual - in Fig.~\ref{fig:starvel}.
Again, our primary aim is to quantify the residual structures; and
the reader is recommended to look at Section~A2.4 of \cite{dav07} as
well as Section~A.1 of \cite{hic09} for details about the bulk properties
of the gas and stellar systems themselves.
The residual was created by fitting and subtracting a velocity map for
a simple axisymmetric disk model generated by the code DYSMAL, which
is described in \cite{cre09}.
The code produces a data cube with 2 spatial and 1 velocity axis, from
which a velocity field can be extracted in the same way as for real
data.
The input parameters needed include the mass distribution and
geometrical configuration of the disk.
In this case, the orientation (position angle and inclination) of the
disk were fixed as in Section~\ref{sec:paaxrat};
the systemic velocity, as well as the total mass and its distribution
(described by a Sersic function with the defining parameters Sersic
index $n$ and effective radius $r_e$) were allowed to vary.
These 4 parameters provide a convenient way to generate a velocity
field appropriate for a disk.
Their values, which are in themselves unimportant, were adjusted by
comparing the output velocity
field to that extracted from the data and minimising the difference,
taking into account the noise in each spaxel.
We emphasize that because we have not included the dispersion in the
minimisation and mass estimation, the mass derived does not
necessarily represent
the true dynamical mass in the central region of NGC\,1097.
However, the purpose of this procedure was solely to derive a
best-fitting velocity field, and for that it is sufficient.

The residual of the stellar velocity field shows no coherent
structures above the noise level.
This lack of evidence for non-circular motions implies that the stars
themselves are not inherently associated with the spiral structure.
This is perhaps not surprising given the rather high velocity dispersion of
150\,km\,s$^{-1}$ at radii from 0.5\arcsec\ to the edge of the SINFONI
field \citep{dav07}.
This implies that the kinematics are dominated by random motions
rather than ordered rotation, hence we are in fact seeing
the bulge stars.
In this case, the 3-armed spiral structure seen in the stellar
continuum must be a passive effect, most likely caused by
absorption from the true gaseous arms.
This was a hypothesis already put forward by \cite{pri05}, and one
which we shall examine in more detail in Section~\ref{sec:h2}.

\section{Molecular Gas}
\label{sec:h2}

\subsection{Gas Distribution and Flux Residuals}
\label{sec:fluxres}

\begin{figure}
\epsscale{.80}
\plotone{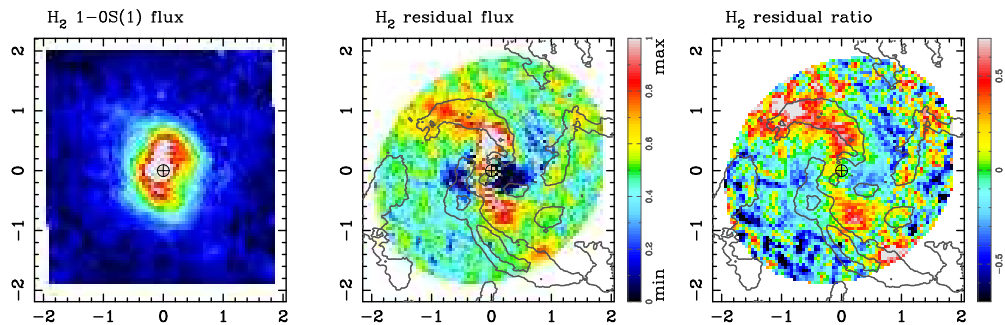}
\caption{
\label{fig:h2flux}
Left: H$_2$ distribution as traced by the 2.12\micron\ 1-0\,S(1) line.
Middle: the flux residual, after subtracting elliptical isophotes,
reveals a 3-armed spiral pattern (2 strong arms and 1 weak) similar to
that seen in stellar absorption. In the flux map, the strength of the
residual increases to within about 0.1\arcsec\ of the centre.
Right: the ratio of the residual flux to the total flux at each spaxel
shows a more uniform distribution along each arm.
For reference, contours outlining the negative J-band stellar residual
are superimposed.
In all panels, north is up and east is left; and the scale in arcsec
  (with 1\arcsec\ = 85\,pc) is indicated.
}
\end{figure}

Within the spectral range for which SINFONI data were obtained, warm
molecular gas at temperatures of 1000-3000\,K can be traced by the
2.12\micron\ 1-0\,S(1) emission line which arises from  H$_2$.
The 2D distribution of this line was extracted as described in
Section~\ref{sec:linefit}, and is presented in Fig.~\ref{fig:h2flux}.
As for the continuum, we have fitted and subtracted elliptical
isophotes to obtain the residual flux distribution. This time,
because the gas is expected to reside in a thin planar disk, we
fixed a common center, position angle, and axis ratio during the fitting
process.

Like the stellar residual flux, presented above, the molecular
gas residuals also reveal an asymmetric distribution of spiral morphology.
Three arms are clearly present, with the northwest one somewhat weaker
and more diffuse. What is immediately apparent is the close spatial
coincidence between the positive H$_2$ residuals and the negative stellar
residuals. This correlation of 1-0\,S(1) H$_2$ emission with
stellar continuum absorption lends strong support to the hypothesis that
the structures in stellar continuum are the result of obscuration.

A closer inspection indicates some spatial offset between the stellar continuum
and molecular flux residuals: the brightest 1-0\,S(1) emission tends to lie
along the inside edge of the arm traced by the J-band stellar residuals. This
offset is clearly seen in the northeastern arm close to the nucleus
(Fig.~\ref{fig:h2flux}), and along the southwestern arm.
It is perhaps to be expected, since the
extinction is roughly proportional to the total amount of gas, while the
1-0\,S(1) emission traces only gas that is heated, probably by shocks, as
it enters the arm along its inside edge (see Section~\ref{sec:models}).
Therefore we refer to the spiral arms traced by the J-band
stellar residual as the `morphological' arms, since we believe they
better represent the true location of gas than the 1-0\,S(1)
emission line.
The picture presented above is complicated about 1\arcsec\ to the southwest
of the nucleus. There the northwestern and southwestern arms appear to
meet, which may be responsible for the shift in the brightest 1-0\,S(1)
emission.

Remarkably, there is an additional hint of the 3-armed spiral
structure in the HCN(1-0) data presented by \cite{koh03} that was
obtained with the Nobeyama Millimeter Array.
Although the resolution was only 4\arcsec$\times$10\arcsec, the
central peak clearly has an extended triangular form, in contrast to the
smoothly elliptical beam.
The vertices of this triangle point in the directions of the
morphological arms on scales of a few hundred parsecs.
An equivalent structure is seen at best only weakly in CO(1-0)
data \citep{koh03,hsi08}, suggesting that the arms are composed
preferentially of denser $n_H>10^4$\,cm$^{-3}$ gas.

It is clear from the H$_2$ residual flux map in the central panel of
Fig.~\ref{fig:h2flux} that as one moves inwards along the arms, the
intensity of the residual 1-0\,S(1) flux increases.
However, this effect disappears if the residual flux is compared to the
total line flux in each spaxel, as shown by the ratio map in the far
right panel.
This suggests that the flux increase in the residuals is related to
the global flux increase at smaller radii.
One possible reason for this could be an increase in the fraction of warm
to cold H$_2$, perhaps due to more efficient excitation at small
radii, for example by the nuclear starburst reported by \cite{dav07} and
\cite{sto05}.
An alternative could simply be an increase in the volume filling
factor of molecular clouds at small radii, an effect that would be
expected given the trend for greater mass surface density at smaller
radii shown in Fig.~7 of \cite{dav07}.

The molecular spiral structure can be traced to within about 0.2\arcsec\ of
the centre, a radius that is formally spatially resolved (the HWHM
from Section~\ref{sec:psf} is 0.1\arcsec).
At this point the residual H$_2$ (in both the flux and ratio maps of
Fig.~\ref{fig:h2flux}) decreases noticeably.
There could be two reasons for this.
The width of the arms is about 0.4\arcsec.
If this size scale is maintained into the centre, one would no longer
expect to discern separate arms closer in than a radius of about
0.2\arcsec.
As a result, any flux closer in than this would therefore be accounted
for during the isophotal fitting and not show up in the residual map.
Alternatively, it could be associated with the increase in the gas
velocity dispersion at about this radius \citep{dav07,hic09}, suggesting
that the gas kinematics are more dominated by random motions, and
hence disk dynamical phenomena are no longer supported.

\subsection{Extinction and gas densities}
\label{sec:gasden}

As we noted above, the spiral structure is present in stellar continuum
images because of a difference in extinction between the arms and the
interarm region. The extinction itself can be derived by comparing the
spectral slope of the K-band continuum in our SINFONI data
to that of stellar templates, as described in \cite{hic09}.
We have used K-type templates, although there is little
difference in the intrinsic slope for any reasonable stellar
population. The low gas velocity dispersion at radii outside 25\,pc
(i.e. the region of interest here) reported by \cite{dav07} and
\cite{hic09} indicates that the gas must lie in a disk, hence the
dust associated with it obscures only half of the stellar bulge light
and a screen approximation is the appropriate model to use for this
extinction.
The resulting average extinction at radii 0.3--2.0\arcsec\ is
$A_V = 4.9\pm1.9$.
If a constant dust to gas ratio is assumed, one can obtain the gas column
density in the disk, which is directly proportional to $A_V$.

\begin{figure}
\epsscale{0.7}
\plotone{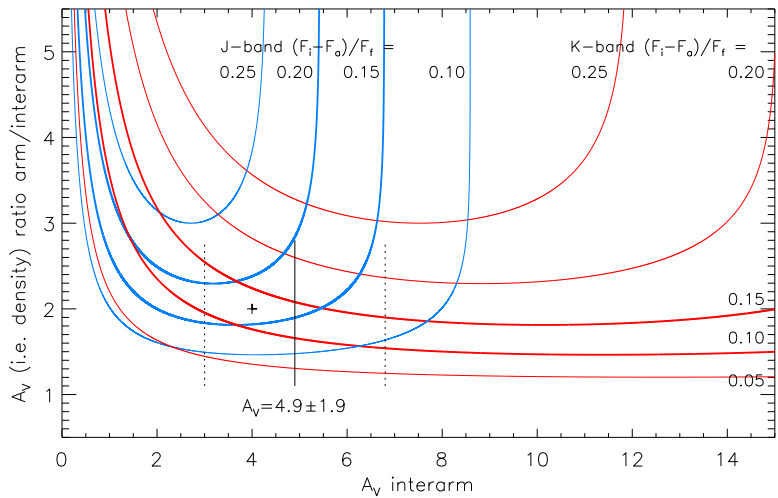}
\caption{
\label{fig:dratio}
Plot of the arm-interarm extinction ratio as a function of interarm
extinction, for several arm-interarm contrast ratios (defined as
$C = F_i/(F_i-F_a)$).
Since $n_H(cm^-2) = 1.9\times10^{22} A_V$ \citep{tok00}, this is
equivalently a plot of the arm-interarm density ratio.
Lines for $C_K$ are drawn in red and $C_J$ in blue;
the thick lines denote the boundaries for the respective measured
values.
The extinction in the central 2\arcsec\ measured from the spectral
slope of the K-band continuum is also overplotted.
Within the uncertainties, the constraints imply an interarm extinction
of $A_V=4\pm1$ and an arm-interarm density ratio of $2.0\pm0.3$.
The adopted value is shown as a plus sign.
}
\end{figure}

Interestingly, one can use the contrast
between the arm and inter-arm regions to relate the interarm extinction to
the extinction in the arm, which gives the arm/interarm gas
density ratio if the dust fraction in the gas is constant.
If we denote the flux intensity on an arm as $F_a$, the inter-arm
flux intensity as $F_i$, and the flux from the isophotal fit as $F_f$,
then one can define a contrast ratio $C$ as the ratio of the difference
between the arm and inter-arm regions to the total flux (as
represented by the fit)
\begin{equation}
C = (F_i-F_a)/F_f.
\end{equation}
As argued above, for dust distributed in the plane of the galaxy, the
appropriate extinction model is that for a screen obscuring only half
the light.
Thus, denoting the intrinsic flux
intensity from the bulge as $F$, we can write
$F_a = 0.5F(1 + x_a)$ and $F_i = 0.5F(1+x_i)$
where $x_a$ and $x_i$ are the respective arm and inter-arm factors by
which the obscured half of the bulge is attenuated, hence the extinction
$A$ is given by
\begin{equation}
A=-2.5\log x.
\label{Ax}
\end{equation}
Since at any point the fitted flux $F_f$ is approximately
the same as the interarm flux $F_i$, we find that
$(x_i-x_a)/(1+x_i) = C$, and hence that there is a linear relation
between $x_a$ and $x_i$:
\begin{equation}
x_a = (1-C)x_i - C.
\end{equation}
This equation, combined with (\ref{Ax}), indicates how the extinction in the
arm depends on the interarm extinction for an observed contrast ratio $C$.
The far right panel in Fig.~\ref{fig:starflux} indicates that the
typical value of $C$ over the central few arcsec in the K-band is
0.10--0.15, i.e. that the K-band continuum is about 10--15\%
fainter on the arms. For $C$ values in this range, the arm/interarm
extinctions should lie in between the two thick red lines in
Fig.~\ref{fig:dratio}.

The contrast ratio observed in our SINFONI K-band data is essentially
the same as in the NACO K-band image presented by \cite{pri05}. In their
J-band data, the typical value of $C$ over the central few arcsec is
0.15--0.20. Since $A_V=3.6A_J=10A_K$, one gets a second set of equivalent
lines for J-band in Fig.~\ref{fig:dratio}. The contrast ratio
in the J-band indicates that the  arm/interarm extinctions should lie in
between the two thick blue lines. Thus in principle, having the contrast
between the arm and interarm regions in more than one band should
enable one to independently determine the extinction and the extinction ratio.
In our case, the combined observed ranges
of the contrast ratio in J-band and K-band confine the possible extinction
and arm/interarm extinction ratio to a narrow range of values. In particular,
these two constraints alone are sufficient to determine the extinction,
without having to know {\em a priori} the intrinsic colours.

If we take the interarm extinction as an approximation for the average
extinction, its value is in excellent agreement with  $A_V = 4.9\pm1.9$
obtained above by independent means. As shown in Fig.~\ref{fig:dratio},
combining the constraints coming from the slope of the spectrum with
those of the contrast ratios, yields a
tight constraint on the extinction and arm-interarm extinction ratio.
Our adopted values are therefore $A_V=4\pm1$, and an
extinction ratio of $2.0\pm$0.3.
Thus the extinction along the arms is $A_V=8.0\pm2.3$.

We note that the extinction we find is higher than the $A_V\sim1$
derived by \cite{pri05}.
The reason is two-fold: these authors assumed that the intrinsic
$J-H$ and $H-K_s$ colours of the bulge were those measured at a
radius of 2.5\arcsec; and they assumed that all the light from the
bulge was extincted.
Correcting for these effects will increase their extinction
significantly, making it more consistent with the value we have
derived.

If the dust fraction in the gas is constant, the arm-interarm extinction ratio
is equal to the gas density ratio, and does not depend on the actual
dust fraction. Thus we estimate the arm-interarm gas density
ratio in the spiral to be $2.0\pm$0.3. On the other hand, in order
to estimate the actual gas column density in the disk, one needs to
rely on the relation between extinction and total hydrogen column
density along the line of sight, $n_H^{sky}$, which for a standard
gas-to-dust ratio of 100 is $n_H^{sky} (cm^{-2}) = 1.9 \times 10^{21}A_V$
\citep{tok00}.
Thus the total projected hydrogen column density between the arms is
$n_H^{sky} = (7.6\pm1.9)\times10^{21}$\,cm$^{-2}$. The corresponding
total deprojected gas surface density in the galaxy disk, $\Sigma_{gas}$,
is expressed by
\begin{equation}
\Sigma_{gas} \, {\rm [M_\odot pc^{-2}]} \; = \; \Sigma_{gas}^{sky} \, \cos i \; =
\; 1.36 \, \Sigma_{H}^{sky} \cos i
\; = \; 10.9 \times 10^{-21} \, \cos i \; n_H^{sky} {\rm[cm^{-2}]},
\end{equation}
where $i$ is the inclination of the obscuring disc, and the 1.36 factor is
the helium correction applied to the total hydrogen density $\Sigma_H$.
In between the arms, we get  $\Sigma_{gas} = 70\pm17$\,M$_\odot$\,pc$^{-2}$.
Along the arms, the corresponding values are $140\pm40$\,M$_\odot$\,pc$^{-2}$.
If we assume that the arms are as thick as they are wide, about
0.4\arcsec, this implies a mean total hydrogen density in the arms of
100--150\,cm$^{-3}$.
Since the gas is likely to exist in clouds rather than being uniformly
distributed, it is consistent with the arms being detected in
HCN(1-0) by \cite{koh03}, which traces gas denser than
$\sim10^4$\,cm$^{-3}$.
These two results are consistent when volume filling factor is no more
than 1--2\%.

Based on the CO(2-1) molecular line luminosity, and the (2-1) to (1-0)
line ratio reported by \cite{hsi08},
we can estimate the average central molecular hydrogen column density
within their observed beam of 3.1\arcsec$\times$4.1\arcsec. Using the
conversion factor of $1.4 \times 10^{20}$ cm$^{-2}$ (K km s$^{-1}$)$^{-1}$
between the (1-0) line intensity and column density of molecular gas, as in
\cite{hic09}, we estimate this density at
$2.8\times 10^{22}$ cm$^{-2}$, corresponding to the
total deprojected gas surface density in the galaxy disk of
260\,M$_\odot$\,pc$^{-2}$. Even if all the gas in this central region were
molecular, this estimate is 3.7 times higher than the interarm
surface density derived above from the stellar continuum extinction.
We discuss the implications of this possible difference in mass surface
density in Section~\ref{sec:disc}.

\subsection{H$_2$ Kinematics and Velocity Residuals}
\label{sec:velres}

\begin{figure}
\epsscale{.95}
\plotone{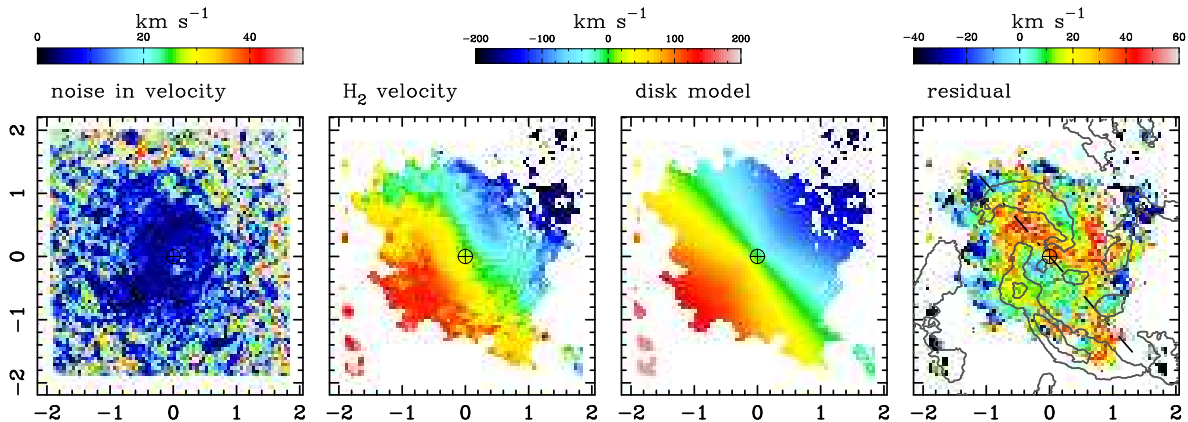}
\caption{Far left: noise in H$_2$ velocity field. The pattern is
  dominated by the effects of the flux distribution (i.e. the
  uncertainty is greatest in the outer regions where the flux is
  lowest).
Center left: the H$_2$ velocity field derived by fitting emission line
  templates to the 2.12\micron\ 1-0\,S(1) line. Regions of particularly
  high noise have been excised.
Center right: the best fitting axisymmetric disk model velocity field.
Far right: the residual after subtracting the model velocity field from
  that observed shows a clear 2-arm spiral pattern.
Intrinsic velocities and residuals are about 50\% higher than shown
  here due to the correction for inclination.
The dashed line in this panel traces the minor axis of the galaxy; and
  the contours outline the negative J-band stellar residual.
In all panels, north is up and east is left; and the scale in arcsec
  (with 1\arcsec\ = 85\,pc) is indicated.
\label{fig:h2vel}
}
\end{figure}

The 2D kinematics of the H$_2$ have been extracted as described in
Section~\ref{sec:linefit} and are presented in Fig.~\ref{fig:h2vel}.
As for the stellar kinematics, we have fitted an axisymmetric disk
model to the velocity field, for a fixed position angle and axis
ratio, allowing the mass and its distribution to vary.
Subtracting this model reveals a clear 2-armed velocity residual.
This 2-arm kinematic spiral pattern winds in the same direction as the 3-arm
morphological pattern seen in the residual flux map, but more tightly.
Understanding how the two patterns are related, and how they
might relate to gas inflow or outflow is developed in Section~\ref{sec:models}.

\begin{figure}
\epsscale{.50}
\plotone{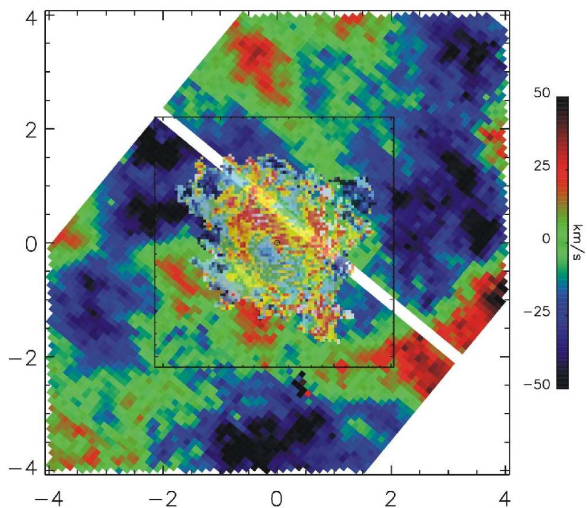}
\caption{Velocity residuals of the [N{\sc ii}] ionised gas from
  \cite{fat06}.
The SINFONI data are superimposed.
Note that \cite{fat06} identified 3 kinematic arms which they
  associated with the morphological arms.
As we argue in the text, the kinematics appear to be consistent with
  those in the 1-0S(1) line, exhibiting a 2-arm
  signature -- traced by positive velocity residual and separated by 2
  regions of negative residual (as explained in the text, the only
  significant exception is immediately west of the nucleus).
In this figure, north is up and east is left; and the scale in arcsec.
\label{fig:fathi}
}
\end{figure}

Intriguingly, a spiral structure in the velocity residual was
already reported by \cite{fat06} based on their analysis of the
[N{\sc ii}] emission line. However, \cite{fat06}
were unable to fully explain their data, probably because
their interpretation was mislead by attempting to
identify a 3-arm spiral in the kinematics:
they argued that there were two `blue-shifted' arms in the residual and one
`red-shifted' arm.
However, the locations of these features do not correspond to the
morphological arms.
We postulate that \cite{fat06} missed one `red-shifted arm', towards the
north, in between their two `blue-shifted' arms. Adding this arm puts
their result in accordance with our 2-arm kinematic spiral: a 2-arm
kinematic spiral has two velocity minima (`blue-shifted' arms) and two
maxima (`red-shifted' arms). It is clear that between two minima a maximum
should exist. Once this missing `red-shifted' arm is added, the kinematic
arms traced by \cite{fat06} in ionised gas kinematic residuals correspond
very well to what we trace in molecular gas kinematic residuals
(Fig.~\ref{fig:fathi}). Two arms marking velocity minima in molecular gas
continue as arms labeled 1 and 3 by \cite{fat06}. One arm marking
velocity maximum that in our data heads towards south-west, continues as
the arm labeled 2 by \cite{fat06}. Another velocity-maximum arm in our
data heads towards the north, where it connects to a region of positive
velocity residual in the ionized gas, which constitutes the arm missed
by \cite{fat06}. Kinematic residuals are less consistent between ionized
and molecular gas in a region about 1\arcsec\ to
the east and slightly north of the nucleus, where we find a positive
residual but \cite{fat06} find a negative residual.
The reason for this may be associated with the map of the [N{\sc ii}] velocity
dispersion in their paper, which shows a region of very high
dispersion at this location.

The amplitude of H$_2$ velocity residuals (40\,km\,s$^{-1}$ before 
correcting for inclination) is similar to that measured by \cite{fat06} 
for [N{\sc ii}] emission. Thus the intrinsic amplitude of radial motions 
observed in the central regions of NGC\,1097, after deprojection, reaches
75\,\kms. Interestingly, it is apparent from Fig.~\ref{fig:h2vel}, that
the most consistent feature of the velocity field in NGC\,1097 is gas 
outflow between the morphological arms, seen within 1 arcsec from the
centre along the minor axis. In NGC\,1097, the velocity
dispersion measured for the warm H$_2$ gas in the central 320\,pc is
about 50\,\kms\ \citep{dav07,hic09}. Since bright 1-0\,S(1) emission is
expected for shock velocities of 20--40\,\kms\ (the molecules
dissociate at higher shock velocity)
this measurement is likely to be an upper limit to the true dispersion
\citep[see][]{hic09}. Applying a quadrature correction yields an intrinsic
dispersion of 30--45\,\kms, which is consistent with the 20--40\,\kms\
[N{\sc ii}] dispersion measured by \cite{fat06}.
This value is also characteristic of the central velocity dispersion
measured in other galaxies from CO emission \citep[e.g.][]{sch06},
which better represents the average state of the dominant gas phase.

\section{Dynamics of Nuclear Spiral Structures}
\label{sec:models}

In the previous sections, we have described how our high spatial
resolution observations of gas morphology and kinematics in the inner
320\,pc of NGC\,1097 recover a spiral pattern that extends to the
innermost 20\,pc of the galaxy.
The pattern is present
both in morphology of the warm gas, traced by the H$_2$ emission, and in the
gas velocity residuals. Below we show that all observed characteristics
of the spiral are consistent with a density wave propagating in a gaseous
disc. In principle, such a density wave can either be self-amplifying, like
in the standard Lin-Shu theory, or driven by a rotating non-axisymmetric
perturbation in the stellar potential. In order for the wave in gas to be
self-amplifying, it has to overcome the stabilizing effect of the stellar
component, in which the spiral structure is apparently absent, and of the
central supermassive black hole. Simple application of the Toomre stability
criterion indicates that the central gaseous disc in NGC\,1097 can become
unstable only when gas surface densities are well above 1000 \solm/pc$^2$
at 100\,pc from the galaxy centre, and still higher at smaller radii.
This is significantly more than the value we have estimated in
Section~\ref{sec:gasden}.
Thus we imply that the nuclear
spiral in NGC\,1097 is a driven density wave, the driver being possibly
one of the bars or another rotating non-axisymmetric perturbation in the
total gravitational potential.

Density waves driven in gas by a rigidly rotating external potential
have been extensively studied \citep[e.g.][]{gol79,mac04a,mac04b}.
In Appendix~A, we summarize the derivation of characteristics of the resulting
spiral pattern within the linear theory. 
We note that the geometry and the amplitude of the spiral does not
depend on the triggering mechanism (Eq.~A8), and the amplitude of
radial velocities is only a weak function of the number of arms (Eq.~A12).
The theory implies (Eq.~A15) that 
the velocity residuals of an $m$-arm spiral density wave should be dominated
by an ($m-1$)-arm kinematic spiral, a correspondence originally noticed 
by \cite{can93}. In NGC\,1097, the
photometric spiral seems to consist of three arms, as originally
noticed by \cite{pri05}, but our observation of the kinematic spiral
(Fig.~\ref{fig:h2vel}) indicates
that this has only two arms, in full agreement with the linear theory. 
Eq.~(A16) shows that within the limits of the linear theory the arms of
the kinematic spiral should be more tightly wound than those of the
morphological spiral. As is apparent from Figs.~\ref{fig:h2flux}
and~\ref{fig:h2vel}, this is also observed in NGC\,1097. Eqs.~(A9)
and~(A10) indicate that the radial velocity is shifted in phase from the
density perturbation by 180\deg\, which means that the density maximum
corresponds to the maximum in inflow velocity, while maximum radial outflow 
is expected at the density minimum (see also the central panel of
Fig.~\ref{fig:appendix}).
The most prominent kinematical feature that we observe in NGC\,1097,
outflow between the spiral arms, is in full agreement with this prediction.

Despite these successes of linear theory in describing what is
observed, we need to be cautious because, as we show below, there are
also some differences between the theory and observations.
Instead, beyond the linear regime, properties of nuclear spirals associated
with a shock wave propagating in gas can be
analysed using the detailed hydrodynamical
models from \cite{mac04b}.
Although these models were constructed for nuclear spiral shocks of 2-arm
morphology and triggered by a bar, a limited quantitative comparison
between the models and NGC\,1097 is possible.

In Eq.~(A12) we show in the linear approximation that the arm/interarm
density amplitude $(1+\epsilon)/(1-\epsilon)$ corresponds to the
amplitude of radial velocity $\epsilon c$, where $\epsilon$ is much
smaller than 1, and $c$ is the isothermal sound speed, or one-dimensional
velocity dispersion in gas. The amplitude of radial motions observed in 
NGC 1097 is significantly larger than the velocity dispersion (75\,\kms\ and
30--40\,\kms, respectively, see Section~\ref{sec:velres}). This indicates that 
the spiral wave observed in NGC 1097 is beyond the linear density wave regime,
and most likely constitutes a shock in gas. Eq.~(A5) shows that in the 
linear regime, the tangent of pitch angle of the spiral wave is 
proportional to the $c/V_{rot}$ ratio, where $V_{rot}$ is the rotational 
velocity. For the central 320\,pc of NGC\,1097 this ratio is about 0.25, 
while tangent of the observed pitch angle ($\sim$60\deg) is much larger, 
about 1.7. The nuclear spiral in NGC\,1097 is too open to be well described
by a linear theory, but both large pitch angle and large amplitude of 
radial velocity can be accounted for when one goes beyond the linear regime,
to the regime when the spiral constitutes a shock in gas.
As discussed above, we can do this by comparison to hydrodynamical models.

\begin{figure*}
\epsscale{1.0}
\plotone{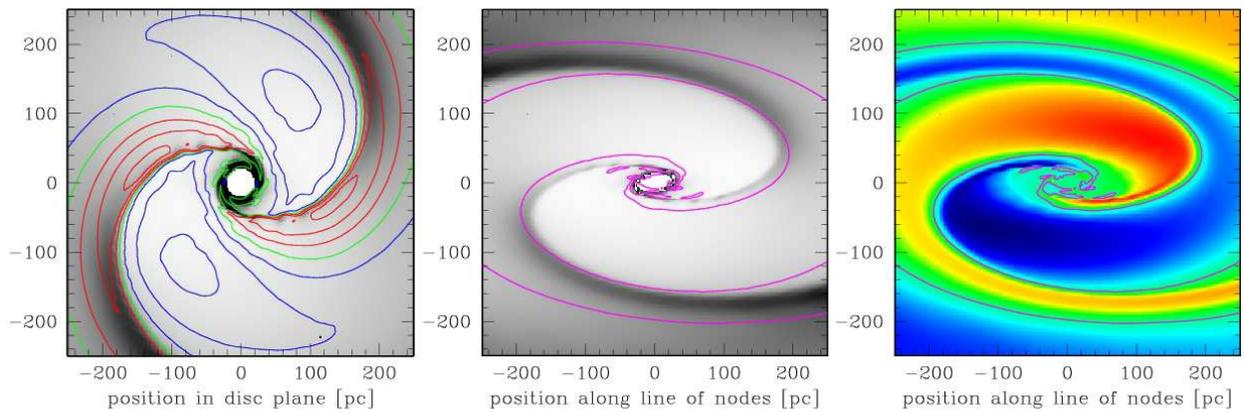}
\caption{
\label{fig:8S20r}
A snapshot of the nuclear spiral shock in gas in model 8S20r from
\cite{mac04b} taken at 300\,Myr, after the flow in the spiral stabilized.
The gaseous disc is rotating clockwise.
{\bf Left:} Radial velocity (contours) plotted over gas density (grey-scale,
darker shades mark higher density).
Blue contours mark positive radial velocity (outflow), green
contours mark zero radial velocity, and red contours mark negative radial
velocity (inflow). The spacing between the contours is 20\,\kms.
The central white hole is outside the inner (reflective) boundary of the
polar grid in the model, hence the model provides no information about this
region, and regions immediately adjacent to it may be influenced by boundary
conditions.
{\bf Centre:} Photometric spiral in gas (grey-scale) as it would appear in
the sky for the galaxy disc inclined at 60\deg. Pink contours mark zero
line-of-sight velocity in gas.
{\bf Right:} Kinematic spiral in the residual gas velocity field. The 
colour scale ranges from -55\,\kms (blue) to 55\,\kms (red). Pink contours are
the same as in the middle panel.}
\end{figure*}

In the left panel of Fig.~\ref{fig:8S20r}, we present the characteristics
of the inner few hundred parsecs of model 8S20r from \cite{mac04b} that
best represents a nuclear spiral associated with a shock. As already
predicted by the linear theory, there is strong inflow in the
morphological spiral arms in the model (red contours), and equally
strong outflow between them (blue contours). \cite{mac04b} identified 
the strong gradient of radial 
velocity on the inside edge of the morphological spiral arm as a
shock in the gas, using negative divergence of the full two-dimensional
velocity field as the shock indicator. The reason for this flow
pattern is that the gas inflowing in the arm preserves some angular
momentum, hence it does not fall onto the galaxy centre, but passes it
by at a certain distance, and continues as a diverging outflow, in which
the gas density decreases. This now low-density, outflowing gas hits the
inside edge of the other spiral arm. This collision maintains the presence
of the shock on the inside edges of the arms. In the shock the gas is again
compressed, and because of losing angular momentum it starts flowing inwards,
so the cycle repeats. Thus the morphological spiral is always downstream
from the spiral shock, which can explain why in NGC\,1097 the photometric
spiral in the stellar residual, coming from obscuration caused by dust
associated with high density gas, is downstream from the spiral in 1-0\,S(1)
emission, which traces warm, recently shocked molecular gas.

The nuclear spiral in the model is loosely wound, which well represents the
morphological spiral observed in NGC 1097. The amplitude of the radial
velocity in the model reaches 60\,\kms, compared to 75\,\kms\ observed in
NGC\,1097 (Section~\ref{sec:velres}). The speed of sound in the isothermal gas
(the one-dimensional velocity dispersion) in the model is 20\,\kms,
while estimates of this value for the centre of NGC 1097 are between
30 and 40\,\kms. Thus the observed Mach number is slightly lower than that in
the model, implying that the observed shock is somewhat weaker than in
the model.
Proper modeling of the nuclear spiral shock would require taking
into account the gas velocity dispersion rising towards galaxy center, which
has not yet been done, and is beyond the scope of this paper.
The ratio of highest density in the
arm to the lowest interarm density in the model can be as high as 10, but
more representative for comparison with the observations is the ratio of
average densities in the arms and between the arms, where arms in the model
are defined as regions of inflow (negative radial velocity in
Fig.~\ref{fig:8S20r}). This ratio is about 2.4, slightly higher than
the arm/interarm density ratio of 2.0 implied by observations (Section~\ref{sec:gasden}).
This again indicates
that the nuclear spiral observed in NGC 1097 hosts a slightly weaker shock
than the one in the model.

The photometric and the kinematic spiral in the model, as seen in the
plane of the sky, are shown in the central and right panels of
Fig.~\ref{fig:8S20r}. A one-arm kinematic spiral in velocity residuals
corresponds to the two-arm photometric spiral. Thus the relation
already found in linear approximation holds also beyond the linear regime,
in a nuclear spiral shock. As we noted above, this relation is also observed
in NGC\,1097. From Fig.~\ref{fig:8S20r} one can also see that the pitch 
angle of the photometric spiral is larger than the pitch angle of the  
kinematic spiral, so this prediction of the linear approximation holds 
also beyond the linear regime. This relation of pitch angles is also 
observed in NGC\,1097 (Section~\ref{sec:velres}).

\section{Gas Transport Rates}
\label{sec:transport}

The rate of radial gas transport $\dot{M}$ can be most simply
estimated from the gas density in the spiral arms, and the radial velocity
as
\begin{equation}
\label{eq:inflow}
\dot{M} = m \; \Sigma \; v_{r} \; \frac{W}{\sin \alpha},
\end{equation}
where $m$ is the number of arms, $\Sigma$ is the mean density of gas in the
arm, $v_{r}$ is the mean radial velocity in the arm, $W$ is the arm width and
$\alpha$ is the pitch angle of the arm. This estimate is valid for rather
loosely wound spirals, with $\alpha \geq$ 45\deg, which is the case of
NGC\,1097 and model 8S20r from \cite{mac04b}.
In NGC\,1097, the mean gas
column density in the arms (see
Section~\ref{sec:gasden}) is 100--180\,\solm/pc$^2$.
The intrinsic mean radial velocity in the arm, after deprojection, is
75\,\kms, and we estimate that the arms are about 0.4\,arcsec (34\,pc)
wide.
The pitch angle of the arms is about 60\deg.
For these values, Eq.~\ref{eq:inflow} gives the radial inflow rate of
$1.23\pm0.35$\,\solm yr$^{-1}$, which is quite significant.

However, already the linear theory indicates that the inflow in the arms
is accompanied by the outflow between the arms (see the middle panel of 
Fig.~\ref{fig:appendix}). Beyond the linear regime, hydrodynamical modelling described in 
Section~\ref{sec:models} shows that at galactocentric radii, where the 
nuclear spiral shock is present, most of the volume is occupied by gas 
that has positive radial velocity component, i.e. is outflowing.
This can be seen in the left panel of Fig.~\ref{fig:8S20r}, where blue
contours that mark positive radial velocities enclose the majority of the
volume. This outflow partially balances the inflow along the arms. However,
in the presence of the shock, the gas on average loses angular momentum
because of its dissipative nature, so there is net inflow in the models. 
For the flow in model 8S20r from \cite{mac04b}, presented in 
Fig.~\ref{fig:8S20r}, this inflow is below 0.01\,\solm yr$^{-1}$ at 
the radius of 100\,pc. On the other hand, the flow parameters in model 
8S20r that enter Eq.~\ref{eq:inflow} are $m=$2,
$\Sigma=$100 \solm/pc$^2$, $v_{r}=$50 \kms, $W=$14 pc, and $\alpha$= 45\deg.
For these values, Eq.~\ref{eq:inflow} gives a radial inflow rate of
0.2\solm yr$^{-1}$. The discrepancy between the actual inflow in the model
and the one estimated from Eq.~\ref{eq:inflow} comes obviously from
the equation not taking into account the interarm outflow.

For NGC\,1097 we cannot derive the inflow directly from the observed data,
and we would like to correct the inflow estimated from Eq.~\ref{eq:inflow} 
for the effects of outflow. The best we can do is to use the ratio of the 
actual to the estimated inflow in model 8S20r as a correction factor, as 
there are no other models that better reflect the observed nuclear spiral
shock. As stated in Section 5, the characteristics of the density wave
do not depend strongly on its driver or the number of arms at least in the
linear regime. Moreover, the kinematics of the nuclear spiral observed in 
NGC 1097 are very similar to those in model 8S20r. With the correction
factor estimated from the model being about 20, the corrected inflow
to the centre of NGC\,1097 should be about 0.06\,\solm yr$^{-1}$. However,
this value is uncertain, as gas flow in the observed 3-arm spiral and
in the modelled 2-arm spiral can differ because of the factors involved
in generating the 3-arm spiral, whose origin remains unclear (see Section
7.4). It is certain, however, that because of the outflow between the arms, 
the net inflow in NGC\,1097 is smaller than $1.2$\,\solm yr$^{-1}$.

\section{Discussion}
\label{sec:disc}

\subsection{Gas Mass and Density}

In order to estimate the average central gas surface density in NGC\,1097,
$\Sigma_{gas}$, we assumed that the gas is distributed uniformly.
On the other hand, the mean volume density we find of
100 -- 150 cm$^{-3}$ is rather less than the $10^4$\,cm$^{-3}$ needed
for bright HCN(1-0) emission.
This implies that the gas distribution, at least in the arms, is most
likely to be clumpy and confined to discrete clouds with a volume
filling factor of only a few percent.
Such dense clouds must be small:
for the inter-arm column density we have estimated, it implies cloud
sizes of no more than 0.2\,pc across.
This suggests that the expression for extinction by a clumpy medium
\citep{cal94} may be more appropriate.
Independent of the actual clouds size, this expression relates the
effective optical depth $\tau_{eff}$ (i.e. that
measured) to the average number of clouds $N$ along the line of sight
and the optical depth through each cloud $\tau_{cl}$ as:
\begin{equation}
\tau_{eff} = N(1-e^{-\tau_{cl}}).
\label{clumpyext}
\end{equation}
For such a distribution, some lines of sight will be relatively little
obscured, while others may be highly obscured.
And if the medium is optically thick on average, one can hide a
considerable column density of gas that has almost no impact on the
observed extinction.
In this regime one would find that $N \tau_{cl} > \tau_{eff}$.
We have used Eq.~\ref{clumpyext} together with the measured extinction
$A_K=0.4$ to estimate the impact of a clumpy medium, and verified the
result statistically.
For $N$ greater than a few (implying cloud sizes $<0.1$\,pc), this
expression yields a quantitative result very similar to that obtained
by assuming a uniform dust screen;
and even in the extreme case of $N=1$, it implies we have
underestimated the column density by only about 20\%.
The reason is simply that in the K-band (i.e. the observational
waveband) individual clouds are not optically
thick, so one is in the regime where $N \tau_{cl} \sim \tau_{eff}$.

The clumpiness of the ISM in NGC\,1097 may help to understand the
different appearances of the VLT NACO J- and K-band images, the
814\,nm HST ACS and 550\,nm HST WFPC2 images (presented in
\citealt{pri05} and \citealt{fat06}).
In the J- and K-band images, the circumnuclear region appears
relatively smooth, although the dust lanes are prominent in the
residual when elliptical isophotes are subtracted.
On the other hand, in the 814\,nm image, a dusty structure is visible
even in the direct image, and appears more like a flocculent spiral.
Then at 550\,nm, the direct image is again relatively smooth.
In terms of the clumpy extinction model, the following argument is
valid as long as there are a few clouds along
the line of sight, and is not dependent on the exact number.
For simplicity let us assume that there are on average $N=4$ along the
line of sight,
so that in the K-band each cloud is optically thin with
$\tau_{cl,K}=0.1$.
On the other hand at 550\,nm even individual clouds will be optically
thick with $\tau_{cl,V}=1$.
Thus in the K-band, the extinction traces the global structure of the
dust through the whole disk.
As a result one sees light from the bulge stars
from in front of the disk and, attenuated, from behind.
In the V-band, nearly every line of sight is fully obscured by the
disk (statistically only 2\% of all lines of sight would not encounter
a cloud).
Thus one sees primarily light from bulge stars in front of the disk,
and the dust structure appears only weakly.
At 814\,nm, one is in an intermediate situation, and hence much more
sensitive to the number of clouds in any particular line of sight.
The structure in the dust distribution therefore shows up more strongly,
and is also influenced by relatively small changes in extinction.
As a result one sees details in the structure as well as the global 3-arm
spiral.
We note that the actual situation may differ from this simplified
picture, since the gas surface density we have derived implies, via the
Kennicutt-Schmidt relation \citep{ken98}, that there is a
low level of on-going star formation in the disk.
However, we believe our simplified analysis is sufficient to
understand what is observed and derive meaningful quantities.

The value for $\Sigma_{gas}$ that we derived in
Section~\ref{sec:gasden}, oscillating
between 140\,M$_\odot$\,pc$^{-2}$ in the spiral arms and
70\,M$_\odot$\,pc$^{-2}$
in the interarm region, has an implication on the gas distribution in the
central few hundred parsecs. The average central gas surface density derived
from the CO(2-1) luminosity is 260\,M$_\odot$\,pc$^{-2}$ over the same region.
Our estimate being lower suggests that the gas may not be uniformly
distributed. In fact, if the gas surface density were as high as the CO data
imply, the flat gaseous disk would extinct 98\% of the J-band light
from the bulge behind it, and we would not see a moderation of that light
caused by obscuration by the nuclear spiral, as we see in
Fig.~\ref{fig:starflux}.
Thus the gas density in the nuclear disk is most likely not as high, as the
CO data imply. On the other hand, if we take the interarm gas surface density
as the average density in the nuclear disk, only 27\% of mass detected in CO
may be distributed in the disk. Since within the CO beam size of
4.1\arcsec$\times$3.1\arcsec, a mass of $1.9\times10^7$\,M$_\odot$ is
implied by the CO emission, only as little as $0.5\times10^7$\,M$_\odot$
would reside in the disk. The rest, $1.4\times10^7$\,M$_\odot$ would be
concentrated in the central few tens of parsecs. \cite{hic09} estimated
the gas fraction within a radius of 1.8\arcsec\ to be $f_{gas}=1.3\pm0.2$\%.
If most of the gas within the central CO beam were to come from an
unresolved nuclear region, then the gas fraction there could be as much as
$f_{gas}\sim15$\%, while being as little as $f_{gas}\sim0.3$\% in the
surrounding nuclear disk.
Only observations at significantly higher resolution than those of
\cite{koh03} and \cite{hsi08} would be able to confirm this conclusion.
However, the low CO intensity in the regions between their central
beams and the circumnuclear ring at a radius of 10\arcsec\ do suggest
that such a low gas fraction is possible on scales of several hundred
parsecs.
It is possible that the explanation for the discrepancy between gas
mass estimated from
extinction and from CO emission may be even simpler. A conversion factor
between the CO emission and the gas column density has to be assumed in
order to obtain the gas masses and surface densities. As pointed out
by \cite{tac08}, around nuclei of galaxies this factor can be up
to a few times smaller than its canonical value. If so, this would put
the CO mass estimates in accordance with our estimates from extinction,
and no gas concentration in an unresolved nuclear region would be required.

\subsection{Inflow Timescales and Nuclear Starbursts}

In Section~\ref{sec:fluxres} we showed that the width of the arms did
not appear to change with radius, and so they could only be
traced in to a radius of about 0.2\arcsec, the point at which the gas
dispersion begins to increase.
This angular distance corresponds to 15--20\,pc, strongly suggesting
that the inflow along the arms is not feeding the AGN directly.
It is on these scales that a nuclear starburst has been found, both by
\cite{sto05} and also by \cite{dav07} who showed in their Fig.~20 that
the Br$\gamma$ flux did appear marginally resolved, covering the
central 0.4--0.5\arcsec.
We conclude therefore that the arms are feeding a gas reservoir in the
central few tens of parsecs, and hence giving rise to starbursts on
these scales.

In NGC\,1097, the data presented by \cite{sto05} and \cite{dav07} are
consistent with a recent starburst creating $10^6$\,M$_\odot$ of
stars in a very short burst that occurred from 1 to 10\,Myr
previously.
For the inflow rate of 0.06\,\solm yr$^{-1}$ that
we estimated in Section~\ref{sec:transport}, it would take 16\,Myr to provide
sufficient gas for 
such a starburst, assuming 100\% star forming efficiency.
For a more realistic efficiency closer to 10\%, the timescale would
increase to order 150\,Myr.
Thus, for a steady-state inflow, one could expect there to be
episodic starbursts, recurring on timescales of order 20--150\,Myr.
For such a scenario, we would currently be in the phase shortly after
a burst.

If the inflow rate estimated in this paper remains unchanged, gas in the
nuclear disk within the central CO beam, whose mass we estimated at
$0.5\times10^7$\,M$_\odot$, is sufficient to last for just five starburst
events. However, \cite{hsi08} estimate the total gas
mass inwards from the nuclear ring, i.e. at radii below 8\arcsec\ at
$2.4\times10^8$\,M$_\odot$,
which, after correcting for the conversion factor between the CO
emission and the gas column density adopted in this paper gives
$1.1\times10^8$\,M$_\odot$. This implies an average gas surface density
of 75\,M$_\odot$\,pc$^{-2}$, virtually the same as our estimate for the
nuclear disk within the central CO beam.
Steady-state inflow in the nuclear spiral of
0.06\,M$_\odot$\,yr$^{-1}$ needs 1.8\,Gyr to drain this reservoir.
It is also possible that this reservoir may be refilled
from the more massive nuclear ring on a similar rate. 
Thus the gas
dynamics inside the nuclear ring implied by the observed morphology
and kinematics, may represent a state that can be sustained for long
timescales.
The nuclear spiral is sufficient to play the role of a mechanism
that feeds gas to the innermost parsecs of a galaxy, where recurrent
star-forming activity takes place. Inflow in the nuclear spiral is up to
two orders of magnitude smaller than in the bar, but it is consistent
with the observed properties of the nuclear stellar population, and represents
a mode of flow that is sustainable for about a Gigayear.

\subsection{Torus}

If the unresolved nuclear region contains a gas reservoir that
supplies recurrent starbursts, an important question concerns
how this gas reservoir relates to the canonical molecular obscuring torus.
This topic has been addressed in detail by \cite{dav06} for NGC\,3227,
\cite{mue09} for NGC\,1068, and by \cite{hic09} for a sample of 9
Seyfert galaxies.
Their conclusion is that the torus is made up of a number of
sub-components that fulfil different roles; and that the molecular gas
concentrations on scales of tens of parsecs are associated with the
outer extent of the overall structure.
This picture is consistent with mid-infrared interferometric data that
have revealed both a bright compact and a more clumpy extended
structures in Circinus \citep{tri07} and NGC\,1068 \citep{rab09};
and also with the simulations performed by
\cite{sch09} (their Fig.~2) which suggest there may be a compact
turbulent disk surrounded by a larger scale component comprising dense
inflowing gas filaments embedded in a hot medium.
Crucially, the structure in the simulations arose as a result of
stellar evolution on scales of tens of parsecs.

In order to try and understand the physical state of the ISM, and how
it could be related to the recurrent starburst episodes, \cite{vol08}
developed a scenario linking these different evolutionary phases
together.
In that paper, the authors argued that a short starburst occurs during
an episode of massive gas inflow to the central few tens of parsecs.
And that during the subsequent supernovae phase, the diffuse
intercloud medium is removed, leaving behind an ISM dominated by
compact dense clumps.
The star formation efficiency in this collisional disk is very low.
The energy to support the scale height is instead supplied by gas
accretion from larger scales; and is redistributed and dissipated via
predominantly elastic cloud collisions, which are in principle
possible if the magnetic fields are sufficiently strong.
As such, the paper addresses the issue of what might be supporting
the large scale height of the disk in the central tens of parsecs.
\cite{hic09} show that this appears to be a common phenomenon in
Seyfert nuclei and
discuss a number of possibilities that might explain it.
They conclude that the AGN itself cannot contribute at the radii
in question; that stellar outflows and supernovae also cannot contribute
significantly; and that stellar radiation pressure could do so, although only
during the active star forming phase.
The alternative of converting gravitational energy of inflowing gas
into turbulence, as suggested by \cite{vol08}, did seem a possible
mechanism as long as cloud collisions were sufficiently elastic.
For the 6 galaxies analysed by these authors, their model predicted
high inflow rates
of several solar masses per year for 2 (including NGC\,1068), and
lower inflow rates of less than 1\,M$_\odot$\,yr$^{-1}$ for the other
4 (among which was NGC\,1097).
Notably, a high inflow rate for NGC\,1068 was also reported by \cite{mue09}
from a detailed analysis of the gas kinematics and distribution in the
central 30\,pc of that galaxy.
For NGC\,1097, the results presented here tend to support the low
0.6\,M$_\odot$\,yr$^{-1}$  accretion rate predicted.
However, when comparing the two values, one should bear in mind that
the inflow rate estimated here is from scales of several hundred
parsecs down to a few tens of parsecs, while that predicted by
\cite{vol08} is from a few tens of parsecs down to a few parsecs.
In addition, we note that 
in the model of \cite{vol08}, the mass inflow rate depends on the gas
mass and its dispersion as
$\dot{M} \propto \sigma^2 M_{gas}^2$.
Thus, if the gas mass is overestimated by a factor of 2,
the required mass accretion rate is reduced by a factor 4.
In this way, moderate uncertainties on these properties can lead to a
large uncertainty on the inflow rate.
Despite the differences and uncertainties, both results point to the same conclusion that the gas inflow rate in the central regions of NGC\,1097 is rather low.

\subsection{Origin of the spiral shock}

The morphology and amplitude of the observed kinematic spiral, presented
in section~\ref{sec:velres} and analyzed in Section~\ref{sec:models}, are
fully consistent with the nuclear spiral in NGC\,1097 being a density wave
in the gaseous disk, associated with a shock, and driven by a rotating
gravitational potential. The driver can be identified with
one of the two bars previously reported in NGC\,1097 (see
\citealt{pri05} for references), or with a possible companion or an
orbiting mass. Nuclear spirals are driven inside the Inner Lindblad
Resonance (ILR) of the driver, if one exists. The analysis of the rotation
curve in NGC\,1097 by \cite{sto96}
indicates that the outer bar definitely has an ILR, hence it can
generate a nuclear spiral. It is possible that the inner bar in NGC\,1097
also has an ILR for the following reason. The rotation curve in NGC\,1097
reaches a plateau at very small radii: about 4-5\arcsec, as implied
from CO and N[II] observations by \cite{hsi08} and \cite{fat06}, and
as small as 1.5\arcsec\ implied by \cite{hic09} from H$_2$ data. For a flat
rotation curve, the ILR appears at 30\% of the corotation radius. The inner
bar in NGC\,1097 ends within the nuclear ring, hence it does not extend
further than 8\arcsec. Numerical models indicate that the corotation of
an inner, nested bar, is located 2-3 times beyond its end (see
\citealt{mac08}, Table 2).
Thus the ILR of the inner bar in NGC\,1097
should occur at 5-7\arcsec, which remains in accordance with the rotation
curve being flat in that region. It is possible therefore that the inner
bar generates its own density wave within the radius of 5-7\arcsec.

\cite{mac04b} constructed hydrodynamical models that enabled him to
study in detail nuclear spirals associated with a shock in gas and
generated by a large-scale bar. In all models 2-arm spirals are generated.
It should be expected, given that the bar is an $m=2$ perturbation in the
potential. However, in NGC\,1097, we observe a 3-arm morphological spiral.
Since another, inner bar is likely present in NGC\,1097, and, as argued above,
it can generate its own spiral pattern, possible non-linear interaction
between the two nested bars in NGC\,1097 may be responsible for the occurrence
of a possibly transient third arm. This effect however has never been
demonstrated in models. To the contrary, recent hydrodynamical models
by \cite{nam09} of gas flows in nested bars report shocks in the secondary
inner bar, but these shocks are still associated with 2-arm spirals.
\cite{nam09} identified one of their models with the dynamics observed in
the centre of NGC\,1097, but, aside for the reason above, that model shows
a different mode of gas flow, because a leading spiral, absent in NGC\,1097,
is present there between the nuclear ring and the central region.

A 3-arm steady-state spiral in the central gaseous disc of a galaxy was
observed only in one type of model, when it was driven by an orbiting
point mass \citep{eth06}. These models were constructed in order to
study possible signatures of remnant black holes that orbit centres of
galaxies.
\cite{eth06} found that a $10^7$\,\solm\ black hole orbiting
at 1-kpc radius generates a 3-arm spiral in the inner few hundred parsecs
of a gaseous disc, but this spiral is weak and tightly wound, unlike the
spiral shock that we observe in NGC\,1097. However, no dependence on gas
velocity dispersion was studied, which in NGC\,1097 is higher than in the
models.

Intriguingly for this scenario, \cite{hig03} showed that the X-shaped
jet-like filaments
extending from the nuclear region of NGC\,1097 to 50\,kpc scales could
be explained if NGC\,1097 had captured a smaller companion in its
central kiloparsecs of order 2\,Gyr previously.
However, it seems unlikely that a small companion could host a black
hole as massive as that required by the models to generate the 3-arm
spiral.

There are claims that the inflow in NGC\,1097 can be of magnetic origin
\citep{bec05}, but they do not attempt to reproduce the observed gas
morphology. Another claim sometimes made about the origin of kinematic
spirals in centres of galaxies in general, is that because of the obscuration
one penetrates to different heights above the disk in the arm and
between the arms, and because rotation velocity decreases with that height,
one sees different velocities in the arm and between the arms, even if in the
disk only axially symmetric rotation is present.
However,
if this was the origin of the kinematic spiral, it would trace the path
of the morphological spiral, would have the same number of arms, and the
residual velocities would be zero on the kinematic minor axis. None
of these three characteristics is true about the kinematic spiral observed
in NGC\,1097, hence the explanation from obscuration has to be rejected.
Interestingly, this also applies to the kinematic spiral observed by
\cite{fat06} in the [N{\sc ii}] emission line, which therefore also cannot
be an effect of the (much greater optical) obscuration.

Thus, given our observations of NGC\,1097, we face an unsettling state of
affairs. The morphology and kinematics of the observed nuclear spiral
have all properties of the spiral shock (velocity residuals, location of
warm molecular gas upstream from the spiral in extinction, $m-1$ multiplicity
of the kinematic spiral, relation of pitch angles of photometric and
kinematic spiral), but we do not know what generates this shock. The
most promising solution is the non-linear interaction of two density
waves generated by the two bars, but more exotic hypotheses, like
orbiting dark matter, cannot be excluded.

\section{Conclusions}
\label{sec:conc}

We have presented near infrared integral field spectroscopic
observations of the central few arcsec (a few hundred parsecs) of
NGC\,1097 at a spatial resolution of 0.2\arcsec\ (20\,pc), focusing
on the distribution and kinematics of the stars and molecular gas.
Our main conclusions are:
\begin{itemize}
\item
We confirm the 3-arm spiral pattern seen previously, and show that
this must be a passive effect on the stellar continuum since there is
no evidence for non-circular motions in the stellar kinematics.

\item
We show that the arms are also traced by H$_2$, and that the shocked
gas lies on the inside edge of the arm. After subtracting an
axisymmetric disk model, the residual velocity field shows a clear
2-arm spiral. Such characteristics are expected if the observed
nuclear spiral is a density wave in gas associated with a shock.

\item
We use the contrast in stellar intensity between the inter-arm and arm
regions, as well as the spectral slope of the stellar continuum, to
estimate the extinction through the galaxy disk.
From this we derive the inter-arm gas mass surface density to be
$\sim70$\,M$_\odot$\,pc$^{-2}$, and that of the arms to
be a factor 2 greater.

\item
A simple estimate of the gas inflow rate along the arms yields
$\sim1.2$\,M$_\odot$\,yr$^{-1}$. However, in a spiral density wave
there is outflow between the arms that partially cancels inflow in the
arms, hence the net inflow rate is smaller than the simple estimate above.
Numerical models indicate that the inflow rate could be 20 times
smaller, i.e. about 0.06\,M$_\odot$\,yr$^{-1}$ in NGC\,1097.
This is sufficient to generate recurrent starbursts, of a scale
comparable to that observed, every 20-150\,Myr.

\item
The nuclear spiral plays the role of a mechanism that feeds gas to the
innermost parsecs of a galaxy, where recurrent episodes of star
formation take place. The inflow rate,
although vastly less than that induced by the large scale bar to the
800\,pc circumnuclear ring, is consistent with the
observed properties of the nuclear stellar population, and represents
a mode of inflow that is sustainable over timescales of
1\,Gyr.

\end{itemize}

\acknowledgments

The authors are grateful to all those at MPE and at Paranal
Observatory who were involved in obtaining these data;
to Almudena Prieto for sharing her NACO imaging data; to
Kambiz Fathi for sharing his [N{\sc ii}] residual velocity data;
and to Bernd Vollmer for interesting and useful discussions.
WM acknowledges an Academic Fellowship from 
Research Councils UK.
Finally, the authors thank the referee for carefully reading the
manuscript and making a number of comments that have helped to improve
it.



{\it Facilities:} \facility{VLT (SINFONI)}



\appendix

\section{Linear approximation for density and l.o.s. velocity
  distribution in spiral density waves driven by a rotating potential}

Density waves in a gaseous disc can be described by analytic solution in
the linear approximation under assumption of tightly wound spiral arms
\citep[for a derivation see Section 6.2.2 of][hereafter BT622]{bin87}.
Usually the solution is given
for self-amplifying waves in self-gravitating gas, but a similar solution
holds for waves driven in a gaseous disc by an external rotating gravitational
potential (e.g. a gaseous disc inside a stellar bar). This solution
has been thoroughly studied \citep{gol79,mac04a},
and here we emphasize its features relevant to the interpretation of the
data presented in this paper.
We follow the notation from \cite{mac04a},
who in turn derived the solution for driven waves by adopting the derivation
for self-amplifying waves given in BT622.

The shape of any arbitrary m-arm spiral can be described in polar
coordinates $(R,\phi)$ by
\begin{equation}
m \phi = f(R).
\label{shape}
\end{equation}
The pitch angle $\alpha$ of this spiral is given by
\begin{equation}
\tan \alpha = \frac{\d R}{R \d \phi} = \frac{m}{R \; \d f/\d R} = \frac{m}{kR},
\label{pitch}
\end{equation}
where the last equality can be written under assumption that the spiral arm
is tightly wound, and then $k \equiv df/dR$ is the radial wavenumber.

The solution for the driven density wave is obtained from the linearized
equations of gas dynamics. It gives the dispersion relation, i.e. the
solution for $k$, which for the driven waves is the same as in the case of
self-amplifying waves (BT622, Eq.~6-40; see also Eq.~16 in \citealt{mac04a}).
After neglecting self-gravity in gas it takes the form
\begin{equation}
k^2(R) c^2 = m^2 (\Omega(R) - \Omega_p)^2 - \kappa^2(R),
\label{k2}
\end{equation}
where $c$ is the sound speed in gas,
$\Omega_p$ is the pattern speed of the driver, $\Omega(R)$ is the angular
velocity in the gaseous disc, and
$\kappa^2 \equiv 2 \Omega (R d\Omega/dR + 2 \Omega)$ is the epicyclic
frequency squared. Having $k$, one can recover the shape of the spiral from
Eq.~(\ref{pitch}). After reshuffling Eq.~(\ref{k2}), the wave number
can be written as
\begin{equation}
k = (1-\frac{\Omega_p}{\Omega}) \eta \frac{m \Omega}{c},
\label{k}
\end{equation}
where $\eta = \sqrt{ 1 - \frac{\kappa^2}{m^2 (\Omega-\Omega_p)^2}}$. Well
inside the corotation of the driver, $\eta$ takes values slightly less than
one for $m\ge3$. This $k$ substituted to Eq.~(\ref{pitch}) gives
\begin{equation}
\tan \alpha = \frac{c}{V_{rot}} \eta^{-1} (1 -\Omega_p/\Omega)^{-1}
\label{tan}
\end{equation}
Thus well inside the corotation of the driver, the pitch angle of the spiral
$\alpha$ is proportional to the ratio of sound speed in gas $c$ to the
rotation velocity $V_{rot}$. This ratio should be small for the linearization
in the tightly wound limit to be applicable.

The linear solution for driven density waves gives also the distribution of
each hydrodynamical variable $X$ (e.g. density $\rho$, velocity) in the
disc plane. In the linear approximation, these variables
are written as the sum of a zeroth-order axisymmetric term $X_0$,
and first-order perturbation $X_1$, which in turn can be expanded in
polar coordinates $(R,\phi)$ as a sum of terms with various multiplicities
$m$, rotating at various rates $\Omega_p$:
\begin{equation}
X_1 (R,\phi,t) = {\rm Re} \sum_{m,\Omega_p} X_a(R) e^{im(\phi-\Omega_p t)}.
\label{1a}
\end{equation}
For driven waves, only terms in Eq.~(\ref{1a}) with $\Omega_p$ equal to the
pattern speed of the driver are excited. For a given gravitational
potential $\Phi_a(R)$ and density distribution $\rho_a(R)$, radial and
tangential velocity components, $v_{Ra}$ and $v_{\phi a}$ are given by
Eqs.~(10) in \cite{mac04a}. Because these equations contain terms with
derivatives, and 
in $H_a = \Phi_a + c^2 \rho_a/\rho_0$, gas density $\rho_a$ rapidly oscillates
with radius $R$, terms with $\rho_a$ in derivative will dominate. Since for
a tightly wound spiral $\d X/\d R \approx ikR$, equations for velocity components
take the form
\begin{equation}
v_{Ra} = -\frac{m (\Omega-\Omega_p)}{k} \frac{\rho_a}{\rho_0},
\hspace{4cm}
v_{\phi a} = -\frac{i \kappa^2}{2 k \Omega} \frac{\rho_a}{\rho_0},
\label{velorig}
\end{equation}
which is the same as that for a self-amplifying wave (BT622, Eq.~6-37),
after neglecting there $\Phi_a$, because the gravitational potential does
not vary
rapidly with radius for driven waves in gas with negligible self-gravity.

The functional form of $\rho_a$ for a spiral wave that has density maxima
at points defined by Eq.~(\ref{shape}) is
\begin{equation}
\rho_a(R) = g(R)e^{-if(R)} = \epsilon \rho_0 e^{-if(R)}
\label{rho}
\end{equation}
where we assumed that the amplitude of the density perturbation $g(R)$
is proportional to the the unperturbed density $\rho_0$ with $\epsilon$
proportionality constant. Substituting Eq.~(\ref{rho}) and
Eq.~(\ref{velorig}) to Eq.~(\ref{1a}) gives first-order perturbation
of gas density and velocity field
\begin{equation}
\rho_1(R,\phi,t) = {\rm Re} \{ \epsilon \rho_0 e^{-if(R) + im (\phi - \Omega_p t)} \}
       = \epsilon \rho_0 \cos (m\phi - f(R) - m \Omega_p t )
\end{equation}
\begin{eqnarray}
v_{R1}(R,\phi,t) = {\rm Re} \{ -\frac{m (\Omega-\Omega_p)}{k} \frac{\rho_a}{\rho_0} e^{im (\phi - \Omega_p t)} \}
       = {\rm Re} \{ -\frac{ m (\Omega-\Omega_p)}{k} \epsilon e^{-if(R) + im
       (\phi - \Omega_p t)} \}
       =  \nonumber\\
-\epsilon \frac{m (\Omega-\Omega_p)}{k} \cos (m\phi - f(R) - m \Omega_p t)
\label{vr1}
\end{eqnarray}
\begin{equation}
v_{\phi 1}(R,\phi,t) = {\rm Re} \{ -\frac{i\kappa^2}{2 k \Omega} \frac{\rho_a}{\rho_0} e^{im (\phi - \Omega_p t)} \}
        = {\rm Re} \{ -\frac{i\kappa^2}{2 k \Omega} \epsilon e^{-if(R) + im (\phi - \Omega_p t)} \}
        = \epsilon \frac{\kappa^2}{2 k \Omega} \sin (m\phi - f(R) - m \Omega_p t)
\label{vp1}
\end{equation}
The term $m\Omega_pt$ indicates rigid rotation with pattern speed of the
driver. We drop it in the remaining part of this Appendix, which corresponds
to presenting the spiral at $t=0$.
Substituting $k$ from Eq.~(\ref{k}) to Eq.~(\ref{vr1}) allows one to
write the radial component of gas velocity as
\begin{equation}
v_{R1} = -\frac{\epsilon c}{\eta} \cos (m\phi - f(R))
\label{vr1e}
\end{equation}
Thus in the linear approximation, if the arm/interarm density ratio in the
spiral is $\frac{1+\epsilon}{1-\epsilon}$, the amplitude of the radial flow
is $\epsilon c$, where $c$ is the speed of sound in gas. This shows maximal
inflow/outflow attainable in the linear flow.

The complete velocity components are the sums of the unperturbed zeroth-order
term, which for the tangential velocity component is the rotation velocity
$V_{rot}$, and the first order perturbation:
\begin{equation}
v_R = 0 + v_{R1} = v_{R1} \hspace{2cm} v_{\phi} = V_{rot} + v_{\phi1}
\end{equation}
The observed line-of-sight velocity is $v_{los} = v_y \sin i$, where $i$ is
the inclination of the disc and
\begin{equation}
v_y = v_R \sin \phi + v_{\phi} \cos \phi
    = V_{rot} \cos \phi - \epsilon [
      \frac{m (\Omega-\Omega_p)}{k}  \cos (m\phi - f(R)) \sin \phi
    - \frac{\kappa^2}{2 k \Omega}  \sin (m\phi - f(R))  \cos \phi]
\label{vysc}
\end{equation}
The same result was obtained by \cite{can97}, who studied density
perturbation in spiral coordinates. Inside the corotation of the driver, the
coefficient at the first component of the sum in brackets is positive,
while the one at the second component is negative. If both coefficients were
of the same magnitude, they would sum up to a single component with angle
dependence $\sin((m-1)\phi-f(R))$. This is the original finding by
\cite{can93} that an $m$-arm density perturbation appears as an
$m-1$-arm kinematic wave.
More accurately, $a \cos \alpha \sin \beta - b \sin \alpha \cos \beta =
\frac{a-b}{2} \sin(\alpha+\beta) - \frac{a+b}{2} \sin(\alpha-\beta)$,
which applied to Eq.~(\ref{vysc}) gives
\begin{equation}
v_y = V_{rot} \cos \phi + \frac{\epsilon}{2k} [
      (m [\Omega-\Omega_p] + \kappa^2/2 \Omega) \sin ([m-1] \phi - f(R)) -
      (m [\Omega-\Omega_p] - \kappa^2/2 \Omega) \sin ([m+1] \phi - f(R)) ]
\label{vypm}
\end{equation}
Thus the $m$-arm photometric spiral, results in a combination of kinematic
spirals with $m-1$ and $m+1$ arms, whose amplitude ratio is $\frac{m(1-\Omega_p/\Omega) + \kappa^2/2\Omega^2}{m(1-\Omega_p/\Omega) - \kappa^2/2\Omega^2}$.
Well within the corotation of the driver this ratio takes values between
$\frac{m+2}{m-2}$ and $\frac{m+\frac{1}{2}}{m-\frac{1}{2}}$, hence the spiral
of $m-1$ arms dominates. The $m-1$ component from Eq.~(\ref{vypm})
corresponds to the spiral shape defined by $(m-1)\phi = f(R)$. 
Because $f(R)$ is the same for
the photometric and the kinematic spiral, Eq.~(\ref{pitch}) implies that
\begin{equation}
\frac{\tan \alpha_{kin}}{\tan \alpha_{phot}}= \frac{m-1}{m}, 
\end{equation}
and the kinematic spiral is more tightly wound, with the ratio of pitch angles
constant inside the corotation of the driver.

\begin{figure*}
\epsscale{1.0}
\plotone{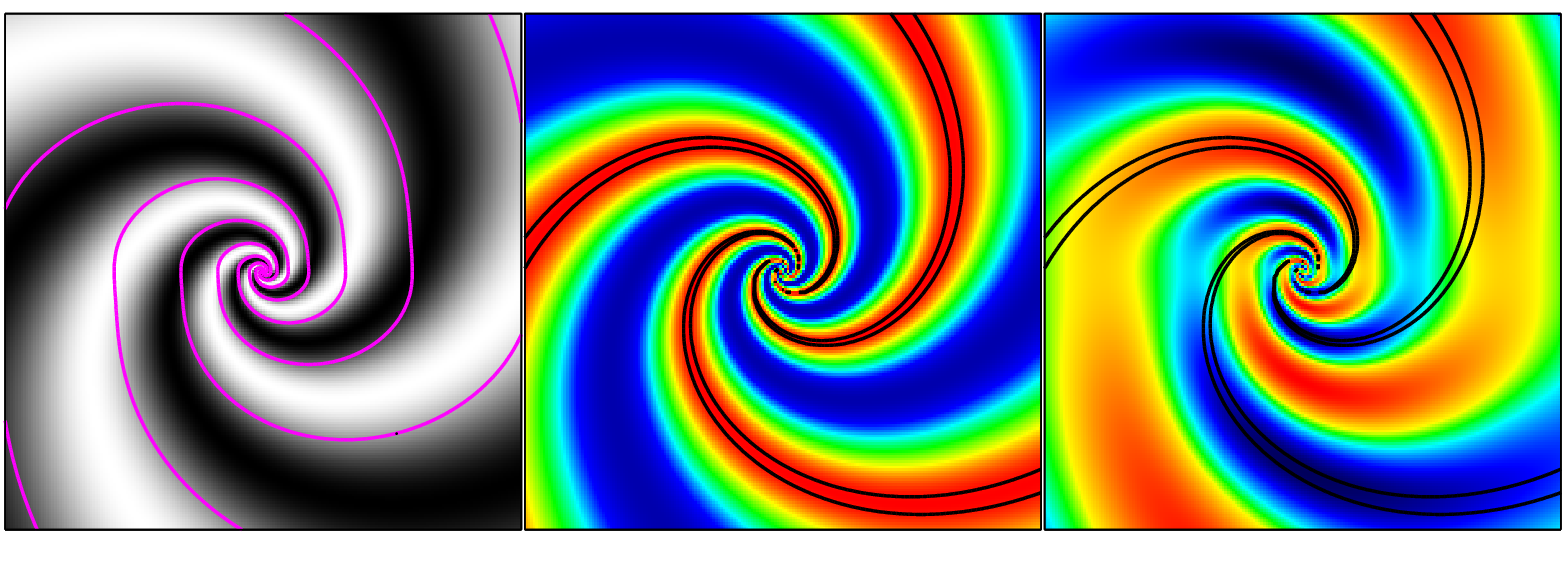}
\caption{
\label{fig:appendix}
An example of a spiral density wave in gas in the linear approximation 
explored in Appendix A. The spiral is shown well inside of the corotation 
of the driver.
{\bf Left:} Morphological 3-arm spiral (grey-scale) with the residual LOS 
(i.e. observed) zero-velocity (pink contours).
{\bf Centre:} Radial velocity in the disc plane (colours: red for inflow, 
blue for outflow), with the density maxima in the spiral arms marked by
black contours. This plot indicates that while morphological spiral arms 
are associated with inflow, there is outflow between the arms in spiral
density waves.
{\bf Right:} The LOS residual velocity field (colours: red for positive, 
blue for negative) after subtraction of circular motions reveals a 2-arm 
kinematic spiral. The density maxima are marked by black contours as in 
the central panel. The line of nodes is horizontal, but the image is 
not projected, as this only involves scaling in the vertical direction.}
\end{figure*}

In Fig.~\ref{fig:appendix}, we visualize the appearance of a 3-arm
photometric spiral, and its corresponding kinematic spiral for
a particularly simple case of a logarithmic spiral, which forms in 
a disk rotating with a constant velocity. In this setup, in the 
region well inside the corotation of the driver ($\Omega \gg \Omega_p$), 
$\eta = \sqrt{1-\frac{2}{m^2}}$ is a constant, and the amplitude 
ratio of the $m-1$-arm and $m+1$-arm kinematic spirals is $\frac{m+1}{m-1}$.
We show the distributions of the density, the radial velocity in the disc
plane and the residual line-of-sight velocity in the innermost 200 pc for 
parameters close to those of the nucleus of NGC 1097: $V_{rot}$ = 
100 km/s, $c$ = 50 km/s and we use $\epsilon$= 0.1.

\clearpage

\end{document}